\documentclass{article}
\pdfoutput=1
\usepackage{spconf,amsmath,graphicx}
\usepackage{xcolor}
\usepackage{multirow}
\usepackage{gensymb}
\usepackage{csquotes}
\usepackage{booktabs}
\usepackage{hyperref}
\usepackage{comment}
\usepackage{url}
\urldef{\ultraxurl}\url{http://www.ultrax-speech.org}

\title{TaL: a synchronised multi-speaker corpus\\of ultrasound tongue imaging, audio, and lip videos}
\name{\parbox{1.0\linewidth}{\center Manuel Sam Ribeiro$^1$, Jennifer Sanger$^1$, Jing-Xuan Zhang$^{1,2}$, Aciel Eshky$^1$,\\Alan Wrench$^{3,4}$, Korin Richmond$^1$, Steve Renals$^1$}
\thanks{Supported by the Carnegie Trust for the Universities of Scotland (Research Incentive Grant number 008585)
and the EPSRC Healthcare Partnerships grant number EP/P02338X/1 (Ultrax2020 –- \ultraxurl).}}

\address{
  $^1$The Centre for Speech Technology Research, University of Edinburgh, UK, \\ 
  $^2${University of Science and Technology of China, P.R.China}, \\
  $^3$Clinical Audiology, Speech and Language Research Centre, Queen Margaret University, UK, \\
  $^4$Articulate Instruments Ltd., UK\\
  {\small \tt \{sam.ribeiro, korin.richmond, s.renals\}@ed.ac.uk}
}

\begin{document}
\ninept

\maketitle

\begin{abstract}
We present the Tongue and Lips corpus (TaL), a multi-speaker corpus of audio, ultrasound tongue imaging, and lip videos. TaL consists of two parts: TaL1 is a set of six recording sessions of one professional voice talent, a male native speaker of English; TaL80 is a set of recording sessions of 81 native speakers of English without voice talent experience. Overall, the corpus contains 24 hours of parallel ultrasound, video, and audio data, of which approximately 13.5 hours are speech. This paper describes the corpus and presents benchmark results for the tasks of speech recognition, speech synthesis (articulatory-to-acoustic mapping), and automatic synchronisation of ultrasound to audio. The TaL corpus is publicly available under the CC  BY-NC  4.0 license.
\end{abstract}

\begin{keywords}
Ultrasound tongue imaging, video lip imaging, silent speech, articulography, corpora
\end{keywords}

\section{Introduction}
\label{sec:intro}

Measuring the position of the articulators during the speech production process is relevant to disciplines such as linguistics, speech processing, speech pathology, and anatomy.
Articulatory movement can be captured  \cite{schultz2017biosignal} using \emph{magnetic articulography techniques}, such as electromagnetic articulography (EMA) or permanent-magnetic articulography (PMA); 
\emph{palatography techniques}, such as electropalatography or optopalatography; or \emph{imaging techniques}, such as video imaging, magnetic resonance imaging (MRI) or ultrasound tongue imaging (UTI).

Video is the cheapest and most convenient imaging method to acquire articulatory data
and is useful for studies focusing on the dynamics of those articulators.
Its main limitation is its inability to capture anything beyond extraoral articulators, such as the lips and jaw.
To provide complementary data, intraoral articulators can be monitored using medical imaging techniques such as MRI \cite{scott2014speech} and UTI \cite{stone2005guide}.
Although MRI captures high-quality images, it suffers from a variety of disadvantages \cite{lawson2015seeing}: 
it is expensive and not easily accessible, and suffers from loud background noise, a supine recording position, and low temporal resolution.
Ultrasound tongue imaging, on the other hand, is portable, non-invasive, clinically safe, and relatively cheap.
UTI uses diagnostic ultrasound operating in B-mode to visualize the tongue surface during speech  production at high frame rates.
There are, however, some challenges associated with ultrasound images of the tongue \cite{ribeiro2019speaker}.
They may be noisy and image resolution may not be ideal, as they are susceptible to unrelated high-contrast edges, speckle noise, or interruptions of the tongue surface. 
Furthermore, image quality may be affected by speaker characteristics such as age or  physiology, or by probe placement.

Ultrasound tongue imaging has been used in various applications, including speech therapy \cite{cleland2015using, cleland2018enabling, sugden2019systematic}, language learning \cite{wilson2006ultrasound, gick2008ultrasound}, phonetics studies \cite{lawson2015seeing}, and the development of silent speech interfaces \cite{denby2010silent}.
Previous work in the context of speech therapy used ultrasound data to develop tongue contour extractors \cite{fabre2015tongue}, animate a tongue model \cite{fabre2017automatic}, and automatically synchronise and process speech therapy recordings \cite{eshky2019synchronising, ribeiro2019ultrasound}.
Additionally, speech recognition \cite{hueber2010development, ji2018updating} and speech synthesis \cite{denby2004speech, csapo2017dnn} from ultrasound images have been used in silent speech interfaces 
to restore spoken communication for users with voice impairments
or to allow silent communication in situations where audible speech is undesirable. 
Ultrasound data has also been used to develop models for 
articulatory-to-acoustic or acoustic-to-articulatory mapping \cite{hueber2011statistical, porras2019dnn}.
Lip videos have been used for research on multimodal speech perception \cite{mcgurk1976hearing}, 
audio-visual synchronisation \cite{chung2016out}, automatic lip reading \cite{fernandez2018survey}, audio-visual speech recognition \cite{afouras2018deep}, and speech reconstruction from silent videos \cite{ephrat2017vid2speech}.

There are relatively few publicly available corpora of ultrasound tongue images. The Silent Speech Challenge dataset \cite{cai2011recognition} consists of approximately 2500 utterances of ultrasound and video images from a single native English speaker.
The UltraSuite repository \cite{eshky2018ultrasuite} contains ultrasound and audio data of 58 typically developing children and 28 children with speech sound disorders in the context of speech therapy.
Because video imaging is easier to acquire, there is a much larger selection of audio-visual datasets of lip videos (see \cite{fernandez2018survey, potamianos2017audio} for recent surveys).
Earlier corpora were concerned with restricted-vocabulary tasks, such as the recognition of the alphabet \cite{matthews2002extraction}, isolated digits  \cite{messer1999xm2vtsdb}, or sentences with limited vocabulary \cite{cooke2006audio}.
Recent work, however, is based on more comprehensive audio-visual datasets 
containing several hundred hours of speech and thousands of speakers \cite{chung2017lip, afouras2018lrs3}.

This paper presents the Tongue and Lips Corpus (TaL), which 
contains synchronised imaging data of extraoral (lips) and intraoral (tongue) articulators from 82 native speakers of English.
Section \ref{sec:tal-corpus} describes the data collection and data post-processing steps of the TaL corpus.
Section \ref{sec:benchmarks} introduces a set of benchmark experiments for automatic speech recognition, speech synthesis, and automatic synchronisation of ultrasound and audio.
We conclude in Section \ref{sec:discussion} with an analysis of speaker performance across tasks and systems.

\section{The TaL Corpus}
\label{sec:tal-corpus}
\label{subsec:data-collection}

\begin{table}[t]
\centering
\caption{TaL corpus at a glance. Gender, accent, and age are self-reported. Accent is categorised here in terms of English (eng), Scottish (sco), or other.}
\label{tab:tal-at-a-glance}
\begin{tabular}{@{}lll@{}}
\toprule
                                            & \textbf{TaL1} & \textbf{TaL80}    \\ \midrule
Number of speakers                          & 1             & 81                \\
Gender (male/female)                        & 1/0           & 36/45             \\
Accent (eng/sco/other)                      & 1/0/0         & 40/36/5           \\
Age (mean/std)                              & 61/-          & 29.4/11.3         \\
Utterances per session (mean/std)           & 263/14        & 205/25            \\ 
Total number of utterances                  & 1582          & 16639             \\ \bottomrule
\end{tabular}%
\end{table}

The TaL corpus (Table \ref{tab:tal-at-a-glance}) comprises
\textbf{TaL1}, a set of six recording sessions of one professional voice talent, a male native speaker of English; 
and \textbf{TaL80}, a set of recording sessions of 81 native speakers of English, without voice talent experience.
All recording sessions took place between October 2019 and March 2020.
Sessions with the experienced voice talent were approximately 120 minutes in duration, and sessions with the remaining speakers were approximately 80 minutes.

Each participant was fitted with the UltraFit stabilising helmet \cite{spreafico2018ultrafit}, which held the video camera and the ultrasound probe.
Data was recorded using the Articulate Assistant Advanced (AAA) software \cite{articulate2010articulate}.
Ultrasound was recorded using Articulate Instruments' Micro system at $\sim$80fps with a 92\degree ~field of view.
A single B-Mode ultrasound frame has 842 echo returns for each of 64 scan lines, giving a $64\times842$ \enquote{raw} ultrasound frame that captures a midsagittal view of the tongue.
Video images of the lips were recorded at $\sim$60fps (greyscale interlaced) and synchronised with audio using Articulate Instruments' SynchBrightUp unit \cite{articulate2010sbu}.
Participants were seated in a hemi-anechoic chamber and audio was captured with a Sennheiser HKH 800 p48 microphone with a 48KHz sampling frequency at 16 bit.

During data collection there was no attempt to normalise the position of the ultrasound probe or video camera across recording sessions.
Instead, we attempted to position the ultrasound probe individually such that the best possible image was captured for each speaker.
Similarly, the video camera was positioned such that the entirety of the lips was visible on screen.
This is a similar process to what we might find in use cases of this technology, such as speech therapy \cite{cleland2015using}.
There were, however, some challenges during data collection, as participants often moved when speaking or the stabilising helmet shifted.
Whenever there was a substantial degradation in image quality, the equipment was manually adjusted, 
and the participants were asked to read a set of calibration utterances.

\begin{table}[t]
\centering
\caption{Recording protocol for TaL80. \enquote{Type} indicates speech type (read or silent speech). \enquote{Shared} indicates that the prompts were read by all speakers in TaL80. The last row does not have a defined number of prompts, as participants read as many as possible until the end of the session.}
\label{tab:recording-protocol}
\resizebox{0.9\columnwidth}{!}{%
\begin{tabular}{@{}lccc@{}}
\toprule
\multicolumn{1}{c}{\textbf{Prompts}}                                  & \textbf{Type} & \textbf{Shared} & \textbf{Count} \\ \midrule
Swallow                                                            & -               & Yes                    & 1              \\
Calibration                                                        & Read            & Yes                    & 2              \\
Rainbow, Accent                                                    & Read            & Yes                    & 24             \\
Calibration                                                        & Silent          & Yes                    & 2              \\
Rainbow (partial)                                                  & Silent          & Yes                    & 6              \\
Harvard                                                            & Silent          & No                     & 9              \\
Harvard                                                            & Read            & No                     & 9              \\
Spontaneous                                                        & -               & No                     & 1              \\ 
TIMIT, VCTK, Librispeech                                           & Read            & No                     & -              \\ \bottomrule

\end{tabular}%
}
\end{table}

Recording prompts were extracted from a variety of sources:
the accent identification paragraph \cite{weinberger2015speech}, the Rainbow Passage \cite{fairbanks1940voice}, the Harvard sentences \cite{institute1969ieee}, the TIMIT corpus \cite{garofolo1993timit}, the VCTK corpus \cite{yamagishi2019cstr}, and the Librispeech corpus \cite{panayotov2015librispeech}. All prompts were spell-checked and adjusted for British English spelling.
Table \ref{tab:recording-protocol} shows a summary of the recording protocol used for each speaker in TaL80.
Participants were asked to swallow and to read two calibration sentences at the beginning and end of each recording session, and before and after any breaks.
The protocol includes a mixture of prompts read by all speakers and prompts read by each speaker.
Some prompts were read silently, audibly, or both.
Additionally, each participant was asked to speak unprompted for 30-60 seconds.
A list of suggested conversation topics was available, although participants were allowed to choose any topic of their preference.
Each TaL1 session follows a similar protocol, with the addition of whispered speech utterances.

\label{subsec:data-preparation}
Using the AAA software, we synchronised the video and ultrasound data to the audio stream.
During the synchronisation process, the video stream was deinterlaced.
We then exported all recorded data from the AAA software.
We normalised waveform levels using the tools available in \cite{itu2010software}.
Using \emph{ffmpeg}, the video data was scaled to $240\times320$ pixels and encoded with \emph{libx264} using \emph{yuv420p} pixel format and a bit rate of 1 Mbps.
Due to a configuration error, the synchronisation signal was corrupted for the first recording session of TaL1 (\emph{day1}).
For this reason, video synchronisation is not available for this session.

For each speaker, utterances were sorted by recording date-time and indexed from 001.
Each filename contains an additional tag, denoting its prompt type.
The tag \textbf{swa} is used for swallows, \textbf{cal} for calibration utterances, \textbf{spo} for spontaneous speech, \textbf{sil} for silent speech, and \textbf{aud} for audible read speech.
The marker \textbf{x} was used with audible and silent tags to indicate cross-prompts (repeated by all speakers).
The absence of this marker indicates that the prompt is unique to the current speaker.
The TaL1 data contains the additional \textbf{whi} tag to indicate whispered speech.
Cross-prompt tags in this data indicate prompts repeated across sessions.
Furthermore,
each utterance consists of five core data types identified by file extension:

\begin{enumerate}
\item The \textbf{prompt} file is identified by extension \emph{.txt} and it includes the prompt text read by the speaker, and the date and time of the recording.

\item The \textbf{waveform}, identified by the file extension \emph{.wav}, is a single-channel RIFF wave file, sampled at 48 KHz with a bit-depth of 16-bit.

\item The \textbf{synchronization signal} is stored identically to the speech waveform, but is identified by the file extension \emph{.sync}. This waveform contains the audio pulses used to synchronise the video and the ultrasound stream.

\item The \textbf{ultrasound} data is stored across two files. Raw ultrasound data is identified by the file extension \emph{.ult}, while the extension \emph{.param} is a text file containing ultrasound metadata (e.g. frames per second, synchronization offset, etc.).

\item The \textbf{video data} is identified by the file extension \emph{.mp4}, which embeds its metadata.
\end{enumerate}

We transcribed all spontaneous speech utterances and included the text transcription in the corresponding prompt file.
Because spontaneous speech utterances can be long in duration (up to 60 seconds), we manually annotated the boundaries of shorter time segments (typically 5-10 seconds).
This segmentation was added as an additional data type for each speaker, identified by the file extension \emph{.lab}.
This additional data type, available only for spontaneous speech utterances, is a text file with the start and end time in seconds for each time segment and their respective transcriptions.

\begin{table}[t]
\centering
\caption{Amount of data for the TaL corpus by prompt type, computed over parallel data streams (audio, video, ultrasound). The \emph{speech} columns estimate data after Voice Activity Detection. All estimates are given in minutes, except for the last row, which is given in hours. Shared data (denoted by \emph{x} in the tag identifier) means that the same set of prompts are read by all speakers. For TaL1, \emph{x} is applied across recording sessions.} 
\label{tab:tal-amount-data}
\resizebox{1.0\columnwidth}{!}{%
\begin{tabular}{@{}llllll@{}}
\toprule

\multicolumn{1}{c}{\multirow{2}{*}{\textbf{Prompt type}}} & \multicolumn{1}{c}{\multirow{2}{*}{\textbf{Tag}}} & \multicolumn{2}{c}{\textbf{TaL1}} & \multicolumn{2}{c}{\textbf{TaL80}} \\ \cmidrule(l){3-6} 
\multicolumn{1}{c}{}                                      & \multicolumn{1}{c}{}                              & \textbf{speech}  & \textbf{total} & \textbf{speech}  & \textbf{total}  \\ \midrule
Read                                               & aud                                               & 53.26            & 88.93          & 501.83           & 829.45          \\
Silent                                             & sil                                               & 0.00             & 6.31           & 0.00             & 62.47           \\
Whispered                                          & whi                                               & 2.98             & 6.02           & -                & -               \\
Read (shared)                                    & xaud                                              & 4.04             & 5.70           & 162.18           & 218.13          \\
Silent (shared)                                  & xsil                                              & 0.00             & 6.37           & 0.00             & 78.19           \\
Whispered (shared)                               & xwhi                                              & 2.62             & 3.83           & -                & -               \\
Spontaneous                                               & spo                                               & 3.68             & 4.74           & 46.50            & 56.24           \\
Calibration                                               & cal                                               & 2.98             & 4.60           & 32.60            & 48.90           \\ \midrule
Total                                                     & -                                                 & 1.16 hrs         & 2.14 hrs       & 12.39 hrs        & 21.90 hrs       \\ \bottomrule
\end{tabular}%
}
\end{table}

The AAA software receives data from each modality separately.
The ultrasound stream begins recording after the video and audio and normally stops recording after.
We release all data streams without trimming them, as they may be useful for a variety of tasks which are not dependent on parallel data.
Identifying parallel data should be trivial, given the available metadata for each data stream.

\begin{figure}[t]
\centerline{\includegraphics[width=8.5cm]{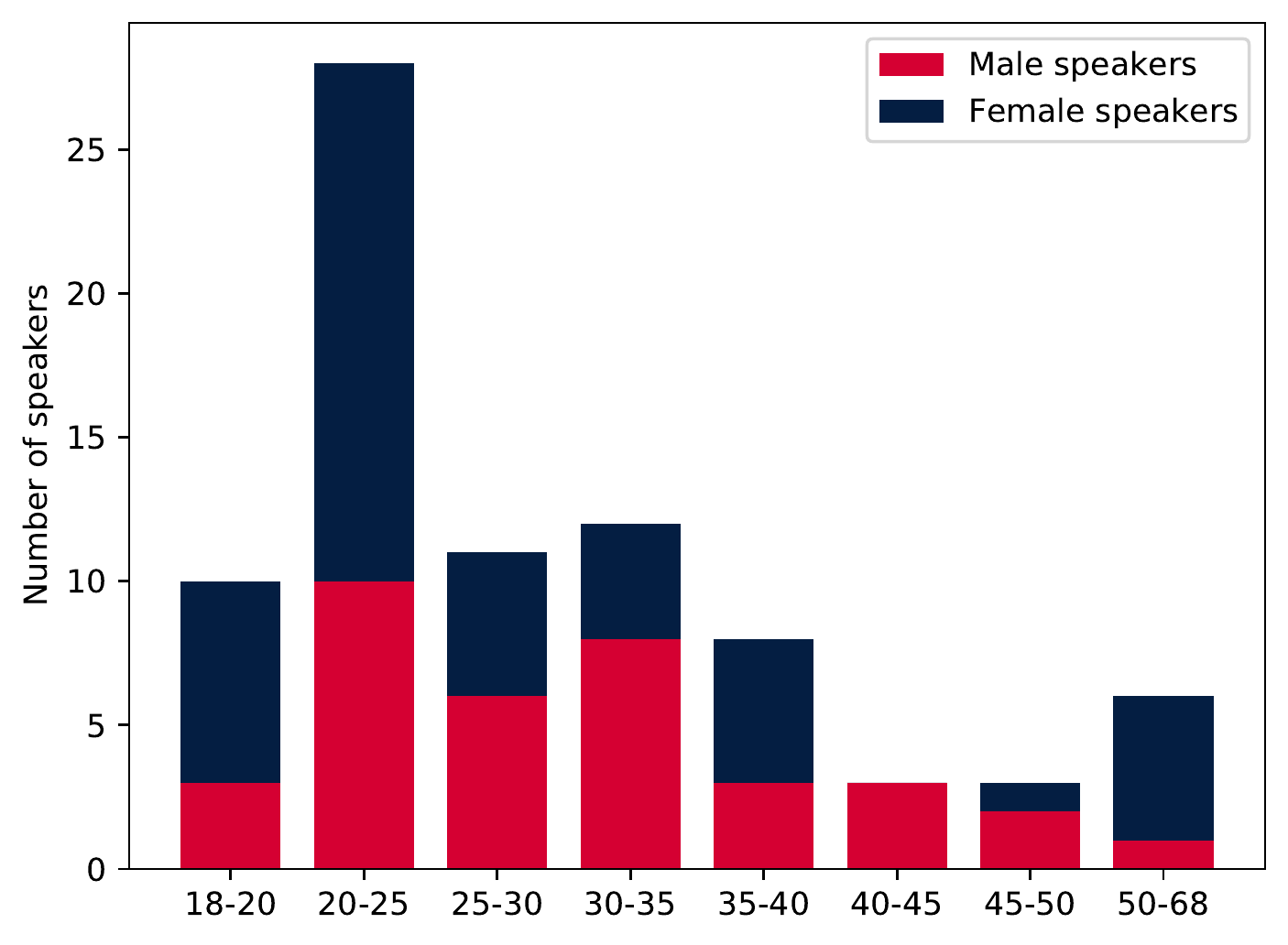}}
\caption{Age distribution for TaL80.}
\label{fig:tal80-ages}
\end{figure}

\begin{figure*}[t]
\centerline{\includegraphics[width=\textwidth]{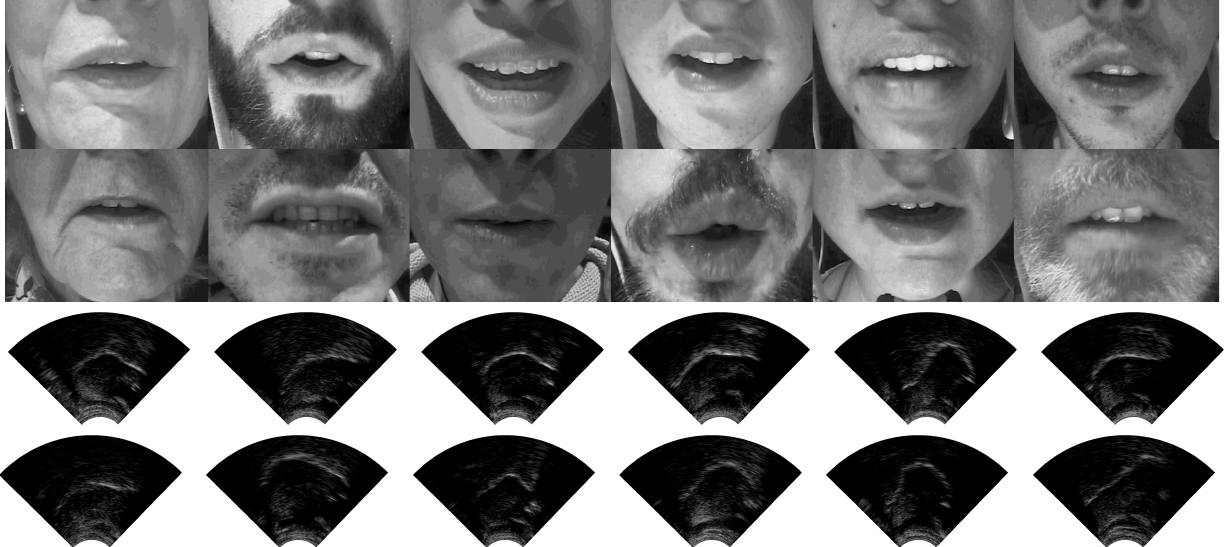}}
\caption{Sample frames from twelve speakers from the TaL corpus. The top figures show video images of the lips and the bottom figures show the corresponding ultrasound images of the tongue.}
\label{fig:tal80-samples}
\end{figure*}

For a summary of the TaL corpus, we refer the reader back to Table \ref{tab:tal-at-a-glance}. 
Table \ref{tab:tal-amount-data} includes more a detailed distribution of the amount of data by prompt.
Overall, we have collected roughly 24 hours of parallel ultrasound, video, and audio, of which approximately 13.5 hours are speech.
This was collected for a total of 82 speakers across 87 recording sessions.
\textbf{TaL1} contains roughly 2 hours of data from a single experienced voice talent across six recording sessions.
\textbf{TaL80} contains almost 22 hours of data from 81 native speakers of English.
Figure \ref{fig:tal80-ages} shows the distribution of ages for TaL80.
There is a slight bias towards younger speakers, as participants were recruited primarily via University channels.
Figure \ref{fig:tal80-samples} shows sample frames from twelve speakers available in TaL80.

\section{Experiments}
\label{sec:benchmarks}
In this section, we present a set of benchmark experiments for the TaL corpus, focusing on the available modalities.
Section \ref{subsec:speech-recognition} presents initial results for automatic speech recognition,
Section \ref{subsec:speech-synthesis} provides results for the articulatory-to-acoustic mapping task, and
Section \ref{subsec:synchronisation} presents results for the automatic synchronisation of ultrasound and audio.

\subsection{Speech recognition}
\label{subsec:speech-recognition}

This set of experiments provides results for speech recognition using the three modalities available in the TaL corpus.
We further consider these results across two training scenarios: speaker dependent and speaker independent.
For this task, we use only read speech utterances (identified by \emph{aud}), and discard silent, spontaneous, whispered, and other shared prompts.
We expect speech styles to have different properties in terms of tongue and lip movement, therefore we focus only on read speech.

For the \textbf{speaker dependent} scenario, we conduct our experiments with the TaL1 dataset.
We discard \emph{day1}, for which video synchronisation is not available.
The remaining sessions, days 2-6, are evaluated using cross-validation, leaving one recording session out for testing.
We aggregate results by computing WER on the total number of errors and words rather than averaging WER across the 5 folds.
For the \textbf{speaker independent} scenario, we train models on speakers 11-81 from the TaL80 dataset.
We discard from the training set all prompts that co-occur with the TaL1 dataset.
We save the first 10 speakers for testing, as these have the biggest prompt overlap with TaL1.
Days 2-6 from TaL1 are used as test set, which is identical to the cross-validation setup for the speaker dependent system.

\emph{Feature extraction}.
During data preparation, calibration utterances are used to identify breaks in the recording session or adjustments to the equipment.
This may indicate different image configurations, so we define a \enquote{speaker} to be any set of utterances without any breaks.
This arrangement allow us to do speaker adaptive training on a speaker dependent scenario.
We extract features over parallel data streams and we trim initial and end silences in the ultrasound and video modalities after Voice Activity Detection on the audio stream.
For the audio stream, we extract 13 dimensional Mel Frequency Cepstral Coefficients (MFCCs) and we append their respective first and second derivatives.
For the visual stream, we apply the 2D Discrete Cosine Transform (DCT) to each ultrasound or video frame.
The upper left submatrix is flattened and used as input.
Video frames are resized to $120 \times 160$ and ultrasound frames are resized to  $64 \times 420$.
Both streams use a submatrix of $12\times12$ coefficients (144 coefficients).

\emph{Model training}.
We train the \enquote{acoustic} models with the Kaldi speech recognition toolkit \cite{povey2011kaldi}.
For these experiments, we initialise models from a flat start using the corresponding features.
After monophone and triphone training, input features are processed with Linear Discriminant Analysis (LDA) and a Maximum Likelihood Linear Transform  (MLLT).
This is followed by Speaker Adaptive Training (SAT) with feature-space MLLR (fMLLR) \cite{rath2013improved}.
The alignment and HMM models from this stage are then used to train a time-delay neural network (TDNN, \cite{peddinti2015time}) following Kaldi's \emph{nnet3} recipe. 
We decode with a simple in-domain bigram language model trained on all prompts from the sources described in section \ref{subsec:data-collection}.

\begin{table}[t]
\centering
\caption{Word Error Rate (WER) for speaker dependent and speaker independent speech recognition on TaL1 and TaL80.}
\label{tab:asr-results}
\resizebox{0.9\columnwidth}{!}{%
\begin{tabular}{@{}cccc@{}}
\toprule
\multirow{2}{*}{\textbf{\begin{tabular}[c]{@{}c@{}}Data\\ streams\end{tabular}}} & \multicolumn{1}{c}{\textbf{Speaker Dependent}} & \multicolumn{2}{c}{\textbf{Speaker Independent}}         \\ \cmidrule(l){2-4} 
                                                                                 & \multicolumn{1}{c}{TaL1}                   & \multicolumn{1}{c}{TaL1} & \multicolumn{1}{c}{TaL80} \\ \midrule
Audio           &  3.42\%   &  4.06\%   &  13.00\%   \\
Tongue          &  38.41\%  &  75.34\%  &  83.81\%   \\
Lips            &  51.57\%  &  84.70\%  &  81.52\%    \\
Tongue+Lips     &  25.09\%  &  49.12\%  &  59.20\%   \\ \bottomrule
\end{tabular}%
}
\end{table}

\emph{Results and future work}.
Results are presented in Table \ref{tab:asr-results}.
As expected, recognition results from the audio stream outperform those of the articulatory modalities by a large margin.
Although we have made no attempts to design more complex systems for silent speech recognition, 
the high WERs illustrate the difficulty of the problem.
This is particularly noticeable when using data from multiple speakers in a speaker independent system.
In all cases, we observe that the visual data streams complement each other well.
The difference in the TaL1 results across the two scenarios shows that speaker independent silent speech recognition is a challenging problem.
Additionally, we observe a large range in the speaker-wise WERs in the TaL80 test set.
Using the Tongue+Lips system, the best speaker achieves a WER of 32.3\% and the worst speaker a WER of 98.3\%.
Further work can focus on identifying speakers with poor image quality and developing methods to improve their results.
There are a number of possibilities that could improve upon these results.
For example, we made no attempt to use the audio stream to initialise the silent models.
Instead of a flat start, we could bootstrap monophone HMMs with audio alignment.
Earlier work showed that substantial improvements can be observed with more complex feature extraction \cite{ji2018updating, tatulli2017feature}.
The results on the TaL80 corpus are speaker independent, but they do make some use of data from target speakers when estimating fMLLR transforms.
We observe that using fMLLR can lead to good improvements.
Other forms of speaker adaptation are likely to be useful.
The multi-speaker systems and fine-tuning results described in the next section further support this idea.

\subsection{Speech synthesis}
\label{subsec:speech-synthesis}

This set of experiments presents results for an articulatory-to-acoustic conversion model.
As in Section \ref{subsec:speech-recognition}, we show results in terms of the three modalities available in the TaL corpus.
Instead of a speaker-independent scenario, we focus here on \textbf{speaker-dependent} and \textbf{multi-speaker} models for speech recovery from silent modalities.
As before, the speaker-dependent model uses days 2-6 from TaL1.
In the multi-speaker condition, we use data from 75 speakers in TaL80 and we hold out the rest for additional experiments.
For training data, we use all the \emph{aud} utterances available for each speaker, of which 10 utterances are randomly held-out for validation.
We use the 24 \emph{xaud} utterances from each speaker for testing.

\emph{Feature extraction}.
The video stream is resampled to match the frame rate of the ultrasound stream using \emph{ffmpeg}.
Each video frame is then resized to $72 \times 136$, randomly flipped horizontally, then cropped to $64 \times 128$ pixels.
Each ultrasound frame is resized to $64 \times 128$.
For acoustic features, we use the STRAIGHT vocoder \cite{Kawahara1999Restructuring} to extract 41-dimensional Mel-cepstral coefficients (MCCs) and 1-dimensional fundamental frequency (F0).
Fundamental frequency is interpolated at the unvoiced frames and converted to log-F0 and a binary voiced/unvoiced flag. 
All features are concatenated to form a 43-dimensional acoustic vector.
Acoustic features are normalized to have zero mean and unit variance for each speaker.
In the multi-speaker model, we additionally use a speaker representation in the form of x-vectors \cite{snyder2018x}, extracted with the Kaldi toolkit.
For waveform reconstruction, we use the STRAIGHT vocoder.

\begin{table}[t]
\caption{Model structure for the articulatory-to-acoustic model.}\smallskip
\centering
\label{tab:syn1}
\begin{tabular}{c l}
\toprule
\multirow{9}{*}{\textbf{Encoder}} & 
3DCNN, $k$ $5\times5\times5$, $s$ $(1, 2, 2)$ , 
 $c$ 32 $\to$ \\
& BN-ReLU-Dropout 0.2 $\to$ \\
& Maxpooling, $k$ $1\times2\times2$, $s$ $(1, 2, 2)$ $\to$ \\
& 3DCNN, $k$ $5\times5\times5$,
$s$ $(1, 2, 2)$,
$c$ 64 $\to$\\
& BN-ReLU-Dropout 0.2 $\to$ \\
& Maxpooling, $k$ $1\times2\times2$, $s$ $(1, 2, 2)$ $\to$\\
& 3DCNN, $k$ $5\times3\times3$,
$s$ $(1, 2, 2)$, 
$c$ 128 $\to$ \\
& BN-ReLU-Dropout 0.2 $\to$ \\
& Flatten $\to$ Fc-512 units $\to$ 
BN-ReLU-Dropout 0.2 \\
\midrule
\multirow{2}{*}{\textbf{Decoder}} & 2-layer BLSTM, 256 units
in each direction $\to$ \\
& Fc-43 units \\
\bottomrule
   \multicolumn{2}{p{220pt}}{$k$ and $s$ represent kernel size and stride respectively. $c$ represents the number of output channels. The dimension order is $T$ (time), $H$ (height), $W$ (width). BN represents batch normalization. Fc represents fully connected layer.}
\end{tabular}

\label{table1}
\end{table}

\emph{Model training}.
We use an encoder-decoder architecture (Table~\ref{tab:syn1}).
The encoder transforms the tongue or lip frames into 512-dimensional vectors.
The decoder then predicts acoustic features conditioned on the encoder outputs and speaker x-vectors.
When both tongue and lip videos are used as inputs, we adopt separate encoders.
The representations from the two encoders are concatenated and sent to the decoder.
For multi-speaker training, the model is first pre-trained on all 75 training speakers and then fine-tuned separately on each speaker.
We used the Adam optimizer with a learning rates of $10^{-3}$ and $10^{-4}$ during pre-training and fine-tuning, respectively.

\emph{Results and future work}.
Results are presented in Table \ref{tab:syn2}.
To evaluate the performance of the model, we report results in terms of mel-cepstral distortion (MCD).
As a proxy measure of intelligibility, we decode the synthesized samples with an open-source ASR model based on ESPnet \cite{espnet} and we report results in terms of WER.
We opt to use this model rather than the audio model described in section \ref{subsec:speech-recognition} as it is not trained on audio data from the target speakers.
Decoding the corresponding natural speech with this model achieves a WER of 0.5\% and 3.5\% for TaL1 and TaL80, respectively.
We observe from Table \ref{tab:syn2} that the performance of the model using different modalities is similar to the results reported in Section \ref{subsec:speech-recognition}.
Using ultrasound tongue images leads to better results than video images of the lips.
The best models use both the ultrasound and video data.
As with speech recognition, the average performance of the multi-speaker system is worse than that of a speaker-dependent system.
We further provide results for fine-tuning the multi-speaker model on the TaL1 data.
This is denoted by \emph{TaL80+TaL1} in Table \ref{tab:syn2}.
Future work will evaluate the intelligibility of synthetic speech with a perceptual test.
To this end, a neural vocoder can be used to generate high quality waveforms.
More powerful encoders, such as ResNet \cite{he2016deep}, can be used to process the video inputs.
As observed in section \ref{subsec:speech-recognition}, results vary substantially across speakers.
Focusing on identifying speakers whose converted speech has low intelligibility and optimizing for their performance could be an interesting future research direction.

\begin{table}[t]
    \centering
    \caption{Mel-Cepstral Distortion (MCD) and Word Error Rate (WER) for the articulatory-to-acoustic conversion model. In the dataset column, TaL1 denotes results for a speaker-dependent model, TaL80 for a multi-speaker model, and TaL1+TaL80 indicates the multi-speaker model fine-tuned on all available TaL1 data. For TaL80, we show mean $\pm$ standard deviation across all speakers.}\smallskip
    \label{tab:syn2}
    \begin{tabular}{c c c c}
    \toprule
    \multirow{2}{*}{\textbf{Dataset}} &
    \multirow{2}{*}{\textbf{\begin{tabular}[c]{@{}c@{}}Data\\ streams\end{tabular}}} &  \multirow{2}{*}{\textbf{MCD (dB)}} & \multirow{2}{*}{\textbf{WER (\%)}} \\
     & \\
    \midrule
    \multirow{3}{*}{\textbf{TaL80}}
    & Lips         & 4.09$\pm$0.19  & 96.0$\pm$3.8   \\
    & Tongue      & 3.35$\pm$0.16  & 56.5$\pm$11.0  \\
    & Tongue+Lips  & 3.31$\pm$0.15  & 53.5$\pm$11.7  \\
    \midrule
    \multirow{3}{*}{\textbf{TaL1}}
    & Lips        & 3.43 & 66.3 \\
    & Tongue     & 2.99 & 27.9 \\
    & Tongue+Lips & 2.84 & 17.2 \\
    \midrule
    \multirow{1}{*}{\textbf{TaL80+TaL1}}
    & Tongue+Lips  & 2.72 & 14.0 \\
    \bottomrule
    \end{tabular}
\end{table}

\begin{table}[t]
\centering
\caption{Accuracy of UltraSync, when pre-trained on out-of-domain child speech therapy data, and when trained on in-domain TaL data.}
\label{tab:sync-results}
\resizebox{0.95\columnwidth}{!}{%
\begin{tabular}{@{}llllll@{}}
\toprule

\multicolumn{1}{c}{\multirow{2}{*}{\textbf{Prompt type}}} & \multicolumn{1}{c}{\multirow{2}{*}{\textbf{Tag}}} & \multicolumn{2}{c}{\textbf{TaL1}} & \multicolumn{2}{c}{\textbf{TaL80}} \\ \cmidrule(l){3-6} 
\multicolumn{1}{c}{}                                      & \multicolumn{1}{c}{}     & \textbf{n} & \textbf{accuracy} & \textbf{n} & \textbf{accuracy} \\

\midrule
\multicolumn{6}{c}{\emph{Out-of-domain model}} \\ \midrule
Read & aud & 384 & 70.6\% & 2595 & 71.6\%\\
Read (shared) & xaud & 48 & 79.2\% & 384 & 72.9\%\\
Spontaneous & spo & 2 & 100\% & 16 & 87.5\%\\
Calibration & cal & 18 & 77.8\% & 134 & 80.6\%\\ \midrule[.02em] 
All & & 452 & 71.9\% & 3129 & 72.3\%\\ \midrule
\multicolumn{6}{c}{\emph{In-domain model}} \\ \midrule
Read & aud & 384 & 98.4\% & 2595 & 97.4\% \\
Read (shared) & xaud & 48 & 97.9\% & 384 & 98.4\% \\
Spontaneous & spo & 2 & 100\% & 16 & 93.8\% \\
Calibration & cal & 18 & 100\% & 134 & 98.5\% \\  \midrule[.02em]
All & & 452 & 98.5\% & 3129 & 97.6\% \\ 
\bottomrule
\end{tabular}%
}
\end{table}

\subsection{Automatic Synchronisation}
\label{subsec:synchronisation}
These experiments present results for the automatic synchronisation of ultrasound and audio. 
The TaL corpus was synchronised at recording time using a hardware synchronisation mechanism which gives the 
offset between the two signals in milliseconds. 
We investigate how well we can predict these offsets by exploiting correlations between the two modalities using the UltraSync architecture \cite{eshky2019synchronising}.

We conduct two experiments using in and out-of-domain data. For the first experiment, we use the UltraSync model from \cite{eshky2019synchronising} which has been trained on \textbf{out-of-domain} child speech therapy data. In the second experiment, we train the UltraSync model on \textbf{in-domain} TaL data. We reserve days 2, 3, and 4 from TaL1 for training, day 5 for validation, and day 1 and 6 for testing. We also reserve speakers 1-49 from TaL80 for training, 50-65 for validation, and 66-81 for testing. We pool the training data and train a single in-domain model from scratch, following the same architecture and training procedures as \cite{eshky2019synchronising}. 
We report all of our results on the same test subset for comparability.

\emph{Data preprocessing}.
We preprocess the data to match the UltraSync input size by resampling the audio to 22.05 KHz using \emph{scipy interpolate}, resampling the ultrasound to 24fps using \emph{skimage transform}, and resizing the ultrasound frames to $63 \times 138$ pixels.

\emph{Experimental setup}.
UltraSync requires us to specify the range of allowable offsets.
In a practical scenario, domain knowledge is utilised to select a suitable candidate list \cite{eshky2019synchronising}.
For these experiments, we consider the observed minimum and maximum offsets given by the hardware sync in the TaL corpus
and we set the step size to 45ms.
We then calculate the range of allowable offsets as:
$[min - (step \times 10),\; max + (step \times 10)]$, which 
renders 25 candidates in total, roughly equal to the 24 candidates in \cite{eshky2019synchronising}. 

\emph{Results and future work}.
For each utterance we calculate the discrepancy:
$disc = prediction - truth$, where $truth$ is the hardware offset. 
A predicted offset is correct if the discrepancy 
falls within the detectability threshold: 
$-$125 $<$ $disc$ $<$ $+$45 \cite{eshky2019synchronising}.
We report results on utterances with audible content and exclude silent and whispered speech in Table~\ref{tab:sync-results}.
The overall out-of-domain model accuracy is 72.2\% 
while the in-domain accuracy is 97.7\%, 
compared to 82.9\% reported by \cite{eshky2019synchronising}.
We attribute the increase in performance
to the higher quality of our data, the absence of multiple speakers in our recordings, and the presence of linguistic variety in our utterances 
compared to child speech therapy data \cite{eshky2019synchronising}.
Future work will explore synchronising the two visual modalities or all three modalities.

\begin{figure}[t]
\centerline{\includegraphics[width=0.95\columnwidth]{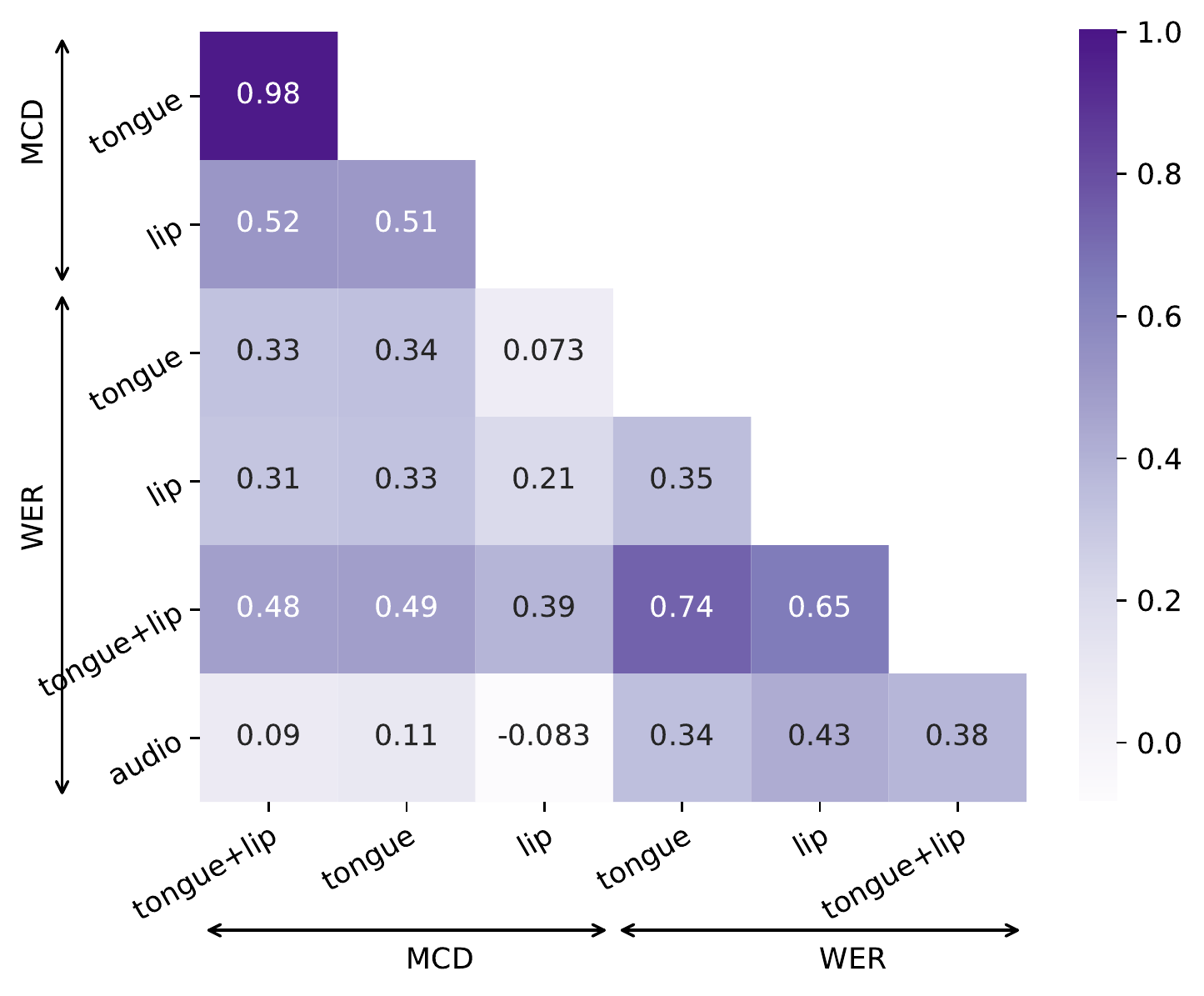}\vspace{-5mm}}
\caption{Pearson's product-moment correlation of speaker-wise results across system pairs. Each cell in this figure indicates the correlation of speaker performance for two systems. WER denotes results from speech recognition systems and MDC denotes results for articulatory-to-acoustic mapping systems.}
\label{fig:speaker-correlation}
\end{figure}

\begin{figure}[ht]
\centerline{\includegraphics[width=0.98\columnwidth]{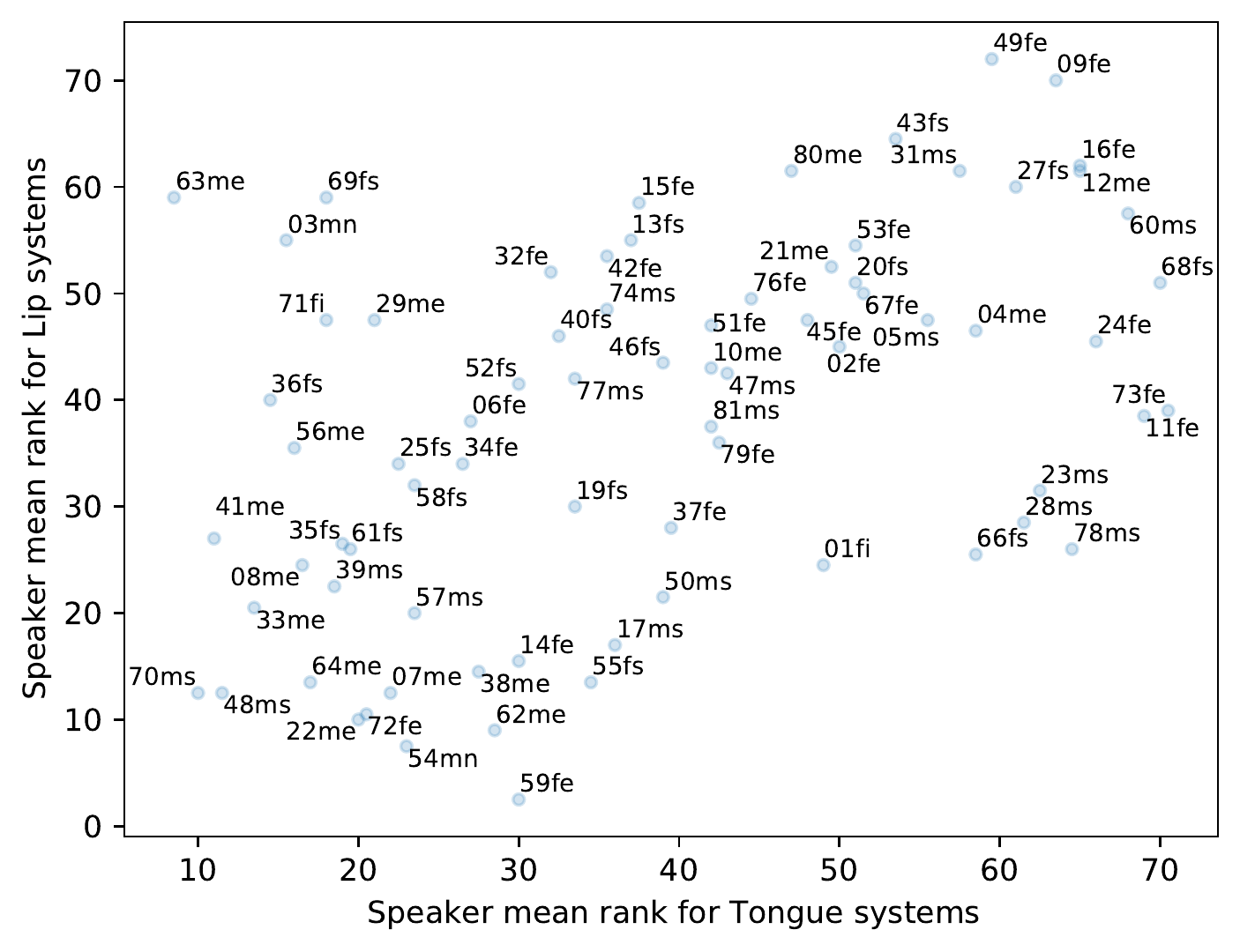}\vspace{-5mm}}
\caption{Speaker mean rank for speech recognition and articulatory-to-acoustic mapping using Tongue-only or Lip-only data for TaL80. Speaker identifiers denote speaker number, gender, and accent.}
\label{fig:speaker-mean-rank}
\end{figure}

\section{Discussion}
\label{sec:discussion}

Experimental results show that results can vary depending on tasks or articulatory modalities.
In this section, we perform an analysis of speaker performance.
We prepare a common evaluation set of all \emph{xaud} utterances available in TaL80 (24 per speaker), which were not seen during training. 
We collect results separately for each speaker using the corresponding speech recognition and articulatory-to-acoustic mapping systems.
We use WER from the TaL80 speaker-independent system in Section \ref{subsec:speech-recognition} and MCD from the TaL80 multi-speaker system of Section \ref{subsec:speech-synthesis}.
Given these speaker-wise WER/MCD scores, we compute Pearson's product-moment correlation across all system pairs (Figure \ref{fig:speaker-correlation}).
We observe that speaker performance is generally more correlated for intra-task systems than inter-task systems, irrespective of modality used.
For example, considering articulatory-to-acoustic mapping (MCD in Figure \ref{fig:speaker-correlation}), speaker performance correlates well  when using different modalities.
That is, speaker-wise MCD results for the systems using tongue-only and lips-only have medium correlation ($r=0.51$).
But when comparing speaker performance across tasks using the same modality, correlation tends to be lower, either with tongue inputs ($r=0.34$) or  lip inputs ($r=0.21$).
We also observe a very high correlation for the articulatory-to-acoustic mapping system using tongue and tongue+lip data ($r=0.98$).
This suggests that the system may not be making the most use of the video data.
This could explain the limited improvements of the tongue+lip system over the tongue system in Section \ref{subsec:speech-synthesis} when compared with the corresponding system in Section \ref{subsec:speech-recognition}.

Additionally, we compare speaker performance on systems using only one of the available modalities.
To find a comparable score across both tasks, each speaker score (WER or MCD) is replaced by their rank order and then averaged across the two tasks.
Figure \ref{fig:speaker-mean-rank} shows speaker mean ranks for systems using only tongue images or lip images.
Although there is low inter-task correlation, Figure \ref{fig:speaker-mean-rank} identifies speakers that underperfom on both modalities and tasks (e.g. \emph{09fe, 16fe, 27fs, 12me}, and \emph{60ms}).
Similarly, some speakers perform well across the two data streams and tasks (e.g. \emph{70ms, 48ms}, and \emph{22me}).
On the other hand, speaker \emph{63me}, for example, achieves good score when using ultrasound data, but scores poorly when using only lip data.
This type of analysis can be used in the future to identify speakers with good-quality imaging data.

\vspace{-0.25mm}
\section{Conclusion}
\label{sec:conclusion}
\vspace{-0.25mm}

We presented the Tongue and Lips corpus (TaL) 
and a set of benchmark experiments for speech recognition, articulatory-to-acoustic mapping, and automatic synchronisation of ultrasound and audio.
These experiments illustrate the challenge of developing robust systems that process articulatory imaging data from multiple speakers.
TaL is released under the Creative Commons Attribution-NonCommercial 4.0 Generic \begin{footnotesize}(CC BY-NC 4.0)\end{footnotesize} licence
and is distributed via the UltraSuite Repository: \url{https://www.ultrax-speech.org/ultrasuite}.

\clearpage 

\bibliographystyle{IEEEbib}
\bibliography{references}

\end{document}